\documentclass[letterpaper,showpacs,aps,prb,floats,twocolumn]{revtex4}
\usepackage{graphicx}
\usepackage{amsmath}
\usepackage{bm,braket}

\usepackage{color}
\usepackage{ulem}
\newcommand {\be}{\begin{equation}}
\newcommand {\ee}{\end{equation}}
\newcommand {\bea}{\begin{eqnarray}}
\newcommand {\eea}{\end{eqnarray}}
\newcommand {\nn}{\nonumber}
\renewcommand{\v}[1]{\ensuremath{\mathbf{#1}}} 

\newcommand{\abs}[1]{\left| #1 \right|} 
\newcommand{\avg}[1]{\left< #1 \right>} 

\def\sm{\sigma^-}
\def\smdag{\sigma^+}
\def\f{\textbf{f}}
\def\fdag{\textbf{f}^\dagger}

\def\r1{\textbf{r}}

\begin{document}

\abovedisplayskip=7pt
\abovedisplayshortskip=7pt
\belowdisplayskip=7pt
\belowdisplayshortskip=7pt

\title{Resonance fluorescence spectra from semiconductor quantum dots coupled to slow-light photonic crystal waveguides}


\author{Kaushik Roy-Choudhury, Nishan Mann, Ross Manson and Stephen Hughes}
\affiliation{Department of Physics, Engineering Physics and Astronomy, Queen's University, Kingston, Ontario, Canada, K7L 3N6}
\email{Corresponding author: kroy@physics.queensu.ca}

\begin{abstract} 
Using a polaron master equation approach we investigate the resonance fluorescence spectra from coherently driven quantum dots (QDs) coupled to an acoustic phonon bath and a photonic crystal waveguide  with a rich local density of photon states (LDOS). Resonance fluorescence spectra from  QDs in semiconductor crystals are known to show strong signatures of electron-phonon interactions, but when  coupled to a structured photonic reservoir, the QD emission properties are also determined by the frequency dependence of the LDOS of the photon reservoir. Here, we investigate the simultaneous role of coupled photon and phonon baths on the characteristic Mollow triplet spectra from a strongly driven QD. As an example structured photonic reservoir, we first study  a photonic crystal coupled cavity waveguide, and find that  photons and phonons have counter-interacting effects near the upper mode-edge of the coupled-cavity waveguide, thus establishing the importance of their separate roles in determining the emission spectra. The general theory is developed for arbitrary photonic reservoirs and is further applied to determine resonance the fluorescence spectra from a realistic, disordered W1 photonic crystal waveguide showing important photon-phonon interaction effects that are directly relevant to  emerging experiments and theoretical proposals.
\end{abstract}


\maketitle

\section{Introduction}

The discrete energy levels of a semiconductor quantum dot (QD) makes it  promising for scalable quantum information applications~\cite{Kim} at optical frequencies. Moreover, the observation of resonance fluorescence spectra~\cite{Mollow} from QDs~\cite{Atature, Ulrich} strongly establishes their atomlike nature mimicking a discrete two-level quantum system when  driven by a coherent laser drive~\cite{Mollow}. Strong field-dressing of the emitter states results in three distinct spectral emissions also known as the Mollow triplet~\cite{Mollow}. Compared to simple two-level atoms, however,  semiconductor QDs are  coupled to the underlying lattice dynamics or phonons which  influence the light emission properties~\cite{Weiler, Forstner, Ramsay, Leonard}. For example, phonon interaction leads to asymmetric line broadening and oscillator strength modification in the Mollow triplet spectra for coherently excited QDs~\cite{Ulrich,Ulhaq}.

Apart from phonon-modified emission dynamics, light-matter interactions in a QD are also influenced by its photonic environment~\cite{Purcell, Dirk},
e.g., resulting in the  Purcell effect or an increase in the spontaneous emission (SE) rate. The SE rate of a quantum emitter is determined by its projected photonic local density of states (LDOS) at the frequency of the emitter; in the domain of strong field resonance fluorescence, the emission spectrum consists of three distinct spectral resonances, which in general depend on a sampling of LDOS values from three different spectral regions of the photonic reservoir~\cite{Ge, John2}, namely at $\omega_L\pm\Omega$, where $\omega_L$ is the frequency of the drive and $\Omega$ is the Rabi frequency. Hence emission from QDs coupled to structured photonic reservoirs can bear signatures of both the coupled  phonon and photon baths. In fact for QD SE decay, interactions between these baths becomes particularly important, when their decay dynamics compare in their time evolution. Emission from QDs under this situation sample a broadband photonic LDOS values around the emitter frequency, in direct violation of Fermi's golden rule~\cite{Kaushik, Kaushik2}. The latter study was  performed under weak driving (excitation) conditions~\cite{Kaushik2,Kaushik3}, when emission is only dominant around the QD frequency. The role of this intercorrelated photon-phonon dynamics is, to teh best of our knowledge,  unknown under strong driving conditions for an arbitrarily shaped LDOS. The problem has been partly investigated, however, in the case of QDs coupled to simple Lorentzian cavities~\cite{Ulrich, Kim2, Arka, Roy}. Coupling with more complex photonic structures (e.g. photonic crystal (PC) band edges~\cite{Roy1}) has been usually treated under a secular approximation~\cite{Roy1, Florescu}, where the calculated side-band intensities do not reflect the underlying asymmetry of the photon bath. Attempts to include correlation effects from simultaneous coupling to structured photon and phonon reservoirs have resorted to simplifying mean-field approximations~\cite{Roy1} for phonons which only leads to a simple overall renormalization of the QD-photon bath coupling. Thus a self-consistent quantum theory describing effects of arbitrary structured photon and phonon reservoirs on QD resonance fluorescence spectra is lacking.

\begin{figure}[th!]
\includegraphics[width=0.95\columnwidth]{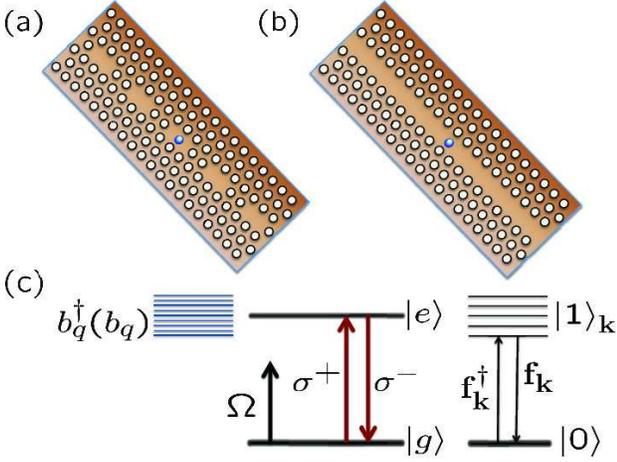}
\centering
\vspace{-0.2cm}
\caption{\label{fig1}(Color online). \footnotesize{  Photonic crystal coupled-cavity waveguide (a) and a (b) W1
PC waveguide (missing row of holes) containing a coupled semiconductor QD.
(c) Energy level diagram of a neutral, coherently driven QD  (electron-hole pair) interacting with a phonon bath and a photon bath. The operator ${\bf f}^\dag_{\bf k}$
($b^{\dag}_q$) creates a photon (phonon).  }}
\label{fig:1}
\vspace{-0.cm}
\end{figure}

In this paper, we develop a polaron master equation (ME) approach~\cite{Roy, Hohenester,  Kaer,Imamoglu}  to  model the resonance fluorescence spectra from a strongly driven QD, simultaneously coupled to an arbitrary structured photon bath and an acoustic phonon bath. We first choose the case of a coupled-cavity PC waveguide (Fig.~\ref{fig1}(a)) which has use for slow-light propagation~\cite{Notomi1} and on-chip single photon emission
 applications~\cite{Rao1,Lodahl,Shields, Laucht}.  The developed model is then applied to investigate the role of photon and phonon bath coupling in the waveguide emitted Mollow triplet spectra. Spectral inhomogeneties of the reservoirs arising due to fast spectral variation of photon and phonon LDOS values gives rise to a Mollow triplet spectrum with asymmetric sideband intensities and linewidths~\cite{Ulhaq}. One consequence of phonon coupling is that   the linewidths of the Mollow sidebands are asymmetric~\cite{Ge} which is also a significant source of line broadening in the QDs. Asymmetric linewidth broadening of Mollow sidebands has been investigated  previously~\cite{Ulrich, Arka, Ulhaq}, but here we focus only on the asymmetry of sideband intensity also in the presence of a rich LDOS photon coupling. The intensity asymmetry arising due to an interplay of  photons and phonons can be clearly distinguished near the upper waveguide mode-edge, where the two baths have counter-interacting effects. The asymmetries of the Mollow spectra as a function of laser-exciton detuning is also investigated to better identify the role of the incoherent photon and phonon scattering mechanisms. The general theory is further applied to study the spectral response of a QD coupled to a disordered W1 waveguide (Fig.~\ref{fig1}(b)), using a realistic LDOS\ model that has been used to explain experiments~\cite{Mann}.

\section{Theory}
We focus on the single exciton state of a neutral QD (strong confinement regime) which can be modelled as a two-level system described by Pauli spin operators $\sigma^{\pm}$ (Fig.~\ref{fig1}(c)). Such a two-level atomic model of a QD has been successfully used to explain numerous experiments investigating phonon-dressed emission from QDs coupled to structured reservoirs~\cite{Ulrich, Weiler, Kim2, Arka}. The QD located at spatial position ${\r1}_d$ is simultaneously coupled to a photon and phonon bath described by lowering operators ${\bf f}(\omega)$ and $b_q$, respectively, where the subscript denotes the $q^{th}$ mode of the phonon reservoir.  The QD is driven by a strong coherent drive with amplitude $\frac{\Omega}{2}$ and frequency $\omega_L$, and the Hamiltonian of the system in the frame of the drive laser is~\cite{Scheel}
\begin{align}
\label{eq1}
 &H = \hbar\int d\v{r} \int_0^{\infty}d\omega\,\fdag(\r1,\omega)\f(\r1,\omega) +
\hbar\Delta_{xL} \smdag\sm \nn\\
&-\left[\smdag e^{i\omega_Lt}\int_0^{\infty}\!d\omega\,\v{d}\cdot\v{E}(\r1_d,\omega) + \text{H.c}\right] + \hbar\frac{\Omega}{2}(\smdag+\sm) \nn\\
&+\Sigma_q\hbar \omega_q b^{\dag}_qb_q  + \smdag\sm\Sigma_q\hbar\lambda_q(b^{\dag}_q+b_q),
\end{align}

\noindent where $\Delta_{xL}=\omega_x-\omega_L$ is the exciton-laser detuning,  $\v{d}=d\v{\hat n}_d$ is the QD dipole moment and $\lambda_q$ is the exciton-phonon coupling strength. The QD interaction with the photonic reservoir is written using the dipole and rotating wave approximation and the electric field operator $\v{E}(\r1,\omega)$ is expressed in terms of the Green's function  $\v{G}(\r1,\r1';\omega)$~\cite{Scheel} of the photonic medium. The QD-phonon interaction term can be eliminated by making a unitary polaron transformation of the Hamiltonian $H$ given by $H' \rightarrow e^P He^{-P}$ where $P = \smdag\sm \Sigma_q\frac{\lambda_q}{\omega_q} (b^{\dag}_q-b_q)$~\cite{Imamoglu}. The polaron-transfromed Hamiltonian $H'$ is given by a sum of three separate contributions:
\begin{align}
\label{eq2}
H' =& H'_{\text{S}}+H'_{\text{R}}+H'_{\text{I}}\nn,\\
H'_{\text{S}}=& \hbar\frac{\Omega_R}{2} [\sigma^++\sigma^-]+\Delta_{xL}\smdag\sm\nn,\\
H'_{\text{R}} =& \hbar\int d\v{r} \int_0^{\infty}d\omega\,\fdag(\r1,\omega)\f(\r1,\omega)+\Sigma_q\hbar \omega_qb^{\dag}_qb_q\nn,\\
H'_{\text{I}} =&   -[B_+\smdag e^{i\omega_Lt}\int_0^{\infty}d\omega\,\v{d}\cdot\v{E}(\r1_d,\omega) + \text{H.c}]\nn\\
&+X_g\zeta_g +X_u\zeta_u, 
\end{align}
 where $X_g = \hbar\frac{\Omega}{2}(\sm+\smdag)$,  
 $X_u = i\hbar\frac{\Omega}{2}(\smdag-\sm)$, and the phonon fluctuation operators~\cite{Imamoglu} are $\zeta_g = \frac{1}{2}(B_+ +B_-- 2\avg{B})$ and $\zeta_u =\frac{1}{2i}(B_+-B_-)$, and $B_{\pm}$ is the phonon bath displacement operator~\cite{Imamoglu}. The polaron modified Rabi frequency of the drive is $\Omega_R = \Omega\avg{B}$, where $\avg{B}$  is the thermal average of $B_{\pm}$ and can be expressed in terms of the independent boson model (IBM) phase $\phi$ as $\avg{B} = \exp[-\frac{1}{2}\phi(0)]$, where $\phi(t) = \int_0^{\infty} d\omega\frac{J_{\rm pn}
(\omega)}{\omega^2}[\coth(\hbar\omega/2k_BT)\cos(\omega t)- i\sin(\omega t)]$ and  $J_{\rm pn}(\omega)$  the phonon spectral function~\cite{Roy3}. For convenience, a polaron shift $\Delta_P= \int_0^{\infty}d\omega\frac{J_{\rm pn}(\omega)}{\omega}$ is implicitly included in the definition of $\Delta_{xL}$. 

In order to derive a polaron ME, we transform the Hamiltonian $H'$ to the interaction picture as $\tilde{H'} \rightarrow U^{\dag}(t)H'U(t)$ where $U(t) = \exp[-i(H'_{\text{S}}+H'_{\text{R}})t/\hbar ]$. A second-order Born approximation in interaction $\tilde{H'_{\text{I}}}$ is the made by assuming a weak coupling between the QD and the photon reservoir and the phonon~\cite{Carmichael} and photon~\cite{Tanas} degrees of freedom are traced out by assuming thermal equilibrium and statistical independence of the two reservoirs~\cite{Carmichael}. The final time-convolutionless~\cite{Breuer} polaron ME, for the QD reduced density operator $\rho$ in the Schr\"{o}dinger picture, is given by~\cite{Kaushik}
\begin{align}
\label{eq3}
\frac{d\rho}{dt} &= \frac{1}{i\hbar}[H'_S,\rho] + \mathcal{L}_{\text{phot}}(\rho) + \mathcal{L}^{\rm D}_{\text{phon}}(\rho) \nn\\&+\gamma_b L(\sm)+\gamma_d L(\smdag\sm),
\end{align}
where the superoperators terms, $\mathcal{L}_{\text{phot}}$ and $\mathcal{L}^{\rm D}_{\text{phon}}$ describe the incoherent interaction of the QD with the photon and phonon bath, respectively, and the Lindblad terms $\gamma_b L(\sm)$ and $\gamma_d L(\smdag\sm)$ describe background SE~\cite{YaoHughes} and pure dephasing~\cite{Borri} of the QD, respectively, where ${L}[O]
=\frac{1}{2}(2O\rho O^\dagger -O^\dagger O \rho-\rho O^\dagger O)$. The term $\mathcal{L}^{\rm D}_{\text{phon}}(\rho)$~\cite{Kaushik} describes phonon modified, incoherent scattering arising due to the coherent drive and is given by,
$\mathcal{L}^{\rm D}_{\text{phon}}(\rho) = -\frac{1}{\hbar^2}\int_0^{\infty} d\tau \sum_{m = g,u}(G_m(\tau)[X_m,e^{-iH'_{\rm S}\tau/\hbar} X_m e^{iH'_{\rm S}\tau/\hbar}\rho(t)]+\text{ H.c.})$, which can be simplified to
an more transparanet analytical form~\cite{Ross} 
\begin{align}
\label{eq4}
\mathcal{L}^{\rm D}_{\text{phon}}(\rho) &= \Gamma^{\sigma^{+}}{L}[\sigma^+]+\Gamma^{\sigma^{-}}{L}[\sigma^-] - \Gamma^{\rm cd}(\smdag\rho\smdag+H.c.)\nn\\
&-(\Gamma_u(\smdag\sm\rho(\smdag-\sm)+\sm\rho) +{\rm H.c.}),
\end{align}
 where the relevant analytical phonon-mediated scattering rates are given by
\begin{align}
\label{eq5}
\Gamma^{\smdag/\sm} &= \frac{\Omega_R^2}{2}\int_0^{\infty} ({\rm Re}[(\cosh{(\phi(\tau))} - 1)f(\tau) \nn\\&+ \sinh{(\phi(\tau))}\cos{(\eta \tau)}] 
\nn\\ &
\mp {\rm Im}[(e^{\phi(\tau)}-1)\frac{\Delta_{Lx}\sin{(\eta \tau)}}{\eta}]d\tau, \nn\\
\Gamma^{\rm cd} &= \frac{\Omega_R^2}{2}\int^{\infty}_0 {\rm Re}[\sinh{(\phi(\tau))}\cos{(\eta \tau)} \nn\\&- (\cosh{(\phi(\tau))}-1)f(\tau)]d\tau, \nn\\
\Gamma_u &= i\frac{\Omega_R^3}{2\eta}\int^{\infty}_0 \sinh{(\phi(\tau))}\sin{(\eta\tau)}d\tau,
\end{align}

\noindent where $f(\tau) = \frac{\Delta_{Lx}^2\cos{(\eta\tau)}+\Omega_R^2}{\eta^2}$ and $\eta = \sqrt{\Omega_R^2+\Delta_{Lx}^2}$. The above rates incorporate the spectral shape of the phonon bath by accounting for phonon damping during Rabi-oscillations of the driven QD and is valid for weak and strong drives~\cite{Ross}. In the limit of weak driving,  previously derived expressions~\cite{Roy3} for the scattering terms $\Gamma^{\smdag/\sm}$ and $\Gamma^{\rm cd}$ are naturally recovered starting from Eq.~(\ref{eq5}). It should also be mentioned that we retain only four relevant terms out of a total of seven analytical phonon rates derived from $\mathcal{L}_{\text{phon}}(\rho)$ in our current work, which uses parameters corresponding to InAs QDs~\cite{Ulhaq}, since the other terms are found to be negligible~\cite{Ross}. Longitudinal acoustic (LA) phonon interaction due to deformation potential coupling plays the strongest role in InAs (and GasAs) QDs, and we use the spectral function $J_{\rm pn}(\omega) = \alpha_p\omega^3 \exp[-\frac{\omega^2}{2\omega_{ b}^2}]$, where the phonon cut-off frequency, $\omega_{ b} = 1$ meV and exciton-phonon coupling strength  $\alpha_{ p}/(2\pi)^2 = 0.06 \rm\, ps^2$~~\cite{Roy}. For these parameters, the derived polaron ME (Eq.~\ref{eq3}) is rigorously valid up to drive strengths of $\Omega~\approx$ 1 meV and this has been explicitly verified (not shown) by matching the results with solutions of weak coupling (i.e, non-polaronic) polaron ME at low temperatures~\cite{Nazir1,Nazir3}.

In addition to the drive-induced phonon scattering rates, incoherent interaction with the photon bath is also influenced by phonons~\cite{Kaushik} and is given by~\cite{Kaushik}
\begin{align}
&\mathcal{L}_{\text{phot}}(\rho) =\! \int_0^t d\tau \!\int_0^{\infty}\!\! d\omega {J_{\text{ph}}(\omega)}[-C_{\text{pn}}(\tau)\smdag\sm(-\tau)e^{i\Delta_L\tau}\rho \nn\\
&+ C^*_{\text{pn}}(\tau)\sm\rho\smdag(-\tau)e^{-i\Delta_L\tau}
+C_{\text{pn}}(\tau)\sm(-\tau)\rho\smdag e^{i\Delta_L\tau} \nn \\  
& -C^*_{\text{pn}}(\tau)\rho\smdag(-\tau)\sm e^{-i\Delta_L\tau}],
\end{align}
 where $\Delta_L = \omega_L-\omega$,  $J_{\text{ph}}(\omega)= \frac{\v{d}\cdot \text{Im}[\v{G}(\r1_d,\r1_d;\omega)]\cdot \v{d}}{\pi\hbar\epsilon_0}$ is the photon reservoir spectral function, and  the {phonon}  correlation function, $C_{\text{pn}}(\tau)=e^{[\phi(\tau)-\phi(0)]}$.
The time-dependent operators $\sigma^{\pm}(-\tau) = e^{-iH'_{\text{S}}\tau/\hbar} \sigma^{\pm}e^{iH'_{\text{S}}\tau/\hbar}$ indicate that the scattering rates are pump-field dependent for strong pumps, and different dressed states ($\omega = \omega_L,\omega_L\pm \Omega_R$) can sample different regions of the photonic LDOS~\cite{Ge,John2}. The incoherent phonon mediated scattering term, $\mathcal{L}_{\text{phot}}(\rho)$ can be simplified using the Markov approximation ($t\rightarrow \infty$) and is given by~\cite{Ge}
\begin{align}
\label{eq6}
&\mathcal{L}_{\text{phot}}(\rho) = \Gamma'(\Omega_R){L}[\sigma]+ [M'(\Omega_R)[\sigma^+,\sigma_z\rho] +H.c]  \nn\\ &-N'(\Omega_R)[\sigma^+\sigma^-,\rho]+ [K(\Omega_R)\sigma^+\rho\sigma^+ + {\rm H.c.}],
\end{align}
 where $\sigma_z =\sigma^+\sigma^--\sigma^-\sigma^+$, and the phonon-modified, drive-dependent scattering rates are 
\begin{align}
\label{eq7}
\Gamma'(\Omega_R) &=2\,{\rm Re}[\frac{\Omega_R^2}{2\eta^2}T_D + \frac{1}{2}(1-\frac{\Omega_R^2}{2\eta^2} -\frac{\Delta_{Lx}}{\eta})T_U \nn\\&+ \frac{1}{2}(1-\frac{\Omega_R^2}{2\eta^2} +\frac{\Delta_{Lx}}{\eta})T_L],\nn\\
N'(\Omega_R) &=i\,{\rm Im}[\frac{\Omega_R^2}{2\eta^2}T_D + \frac{1}{2}(1-\frac{\Omega_R^2}{2\eta^2} -\frac{\Delta_{Lx}}{\eta})T_U \nn\\&+ \frac{1}{2}(1-\frac{\Omega_R^2}{2\eta^2} +\frac{\Delta_{Lx}}{\eta})T_L],\nn\\
M'(\Omega_R) &= \frac{\Omega_R}{2\eta}[\frac{\Delta_{Lx}}{\eta}T_D + \frac{1}{2}(1-\frac{\Delta_{Lx}}{\eta})T_U \nn\\&- \frac{1}{2}(1+\frac{\Delta_{Lx}}{\eta})T_L ],\nn\\
K'(\Omega_R) &=\frac{\Omega_R^2}{2\eta^2}(T_D - \frac{1}{2}(T_U+T_L)),
\end{align}

\noindent where $T_k = \int_0^{\infty}d\tau C_{\rm pn}(\tau)J^k_{\rm ph}(\tau)$, with  the photon relaxation function $J^k_{\text{ph}}(\tau)=\int_0^{\infty}d\omega J_{\text{ph}}(\omega)e^{i(\omega_k-\omega)\tau}$, and $\omega_{k = D,U,L}$ denotes the frequencies of the three dressed states given by $\omega_D=\omega_L$ and $\omega_{U/L} = \omega_L\pm\Omega_R$, respectively. In the absence of phonon coupling (i.e., $C_{\rm pn}(\tau) =1$), then the rates $\Gamma'(\Omega_R),K'(\Omega_R),M'(\Omega_R)$ reduce to the photon reservoir scattering rates $\Gamma(\Omega),K(\Omega),M(\Omega)$ derived  in Ref.~\onlinecite{Ge} and the rates depends solely on the photonic LDOS at $\omega_L,\omega_L\pm\Omega$. However in presence of phonon coupling, a broad range of photonic LDOS around these dressed state frequencies can be sampled by the phonon bath~\cite{Kaushik}, in addition to a polaronic reduction of the Rabi pump field. For weak driving, only photonic LDOS values around the drive frequency $\omega_L$ contributes and $\mathcal{L}_{\text{phot}}(\rho)\rightarrow\frac{\tilde\gamma}{2}L(\sigma)$, where the QD SE rate is simply $\tilde\gamma = 2 {\rm Re}[\int_0^{\infty}d\tau C_{\rm pn}(\tau)J^D_{\rm ph}(\tau)]$~\cite{Kaushik2}. As shown elsewhere, $\tilde\gamma$ violates  Fermi's golden rule due to contribution from broadband photonic LDOS~\cite{Kaushik}. 

The incoherent spectrum from a structured photonic reservoir observed at a detector located at ${\bf r}_{\rm D}$, is given by~\cite{Kaushik3, Ge}
\begin{align}
\label{eq8}
S_P(\r1_{\rm D},\omega) &=  \alpha_{\rm P}({\bf r}_{\rm D},{\bf r}_{d};\omega)\, S_0(\omega),
\end{align}
\noindent where the QD polarization spectrum, 
\begin{align}
\label{eq9}
S_0(\omega) &= \lim_{t\rightarrow\infty}\text{Re}[\int_0^{\infty}d\tau(\avg{\smdag(t+\tau)\sm(t)}e^{\phi(\tau)}\nn\\&-\avg{\smdag\sm})e^{i(\omega_L-\omega)\tau}],
\end{align}
 \noindent and the IBM phase $\phi(\tau)$ appears in $S_0$ due to transformation from a polaron frame to the lab frame~\cite{Kaushik3}, and the propagator, $\alpha_{\rm P}({\bf r}_{\rm D},{\bf r}_{d};\omega)$, of the photon reservoir, describes propagation from the QD  to the detector at  ${\bf r}_{\rm D}$. The projected spectrum $S_{P}$ can be experimentally observed, e.g.,  through the waveguide modes and the polarization spectrum $S_0$ can be observed using the background decay channels $\gamma_b$ (e.g., above light line modes in a PC\ slab waveguide~\cite{YaoHughes}).  All our spectrum calculations make use of the quantum optics toolbox~\cite{Tan}. It should be noted that the polaron ME in its original integro-differential form (Eq.~\ref{eq3}) can of course be numerically solved for a given structured reservoir and we have checked that the calculated spectrum exactly matches the spectrum computed using polaron ME with the incoherent scattering rates (not shown). Naturally, the analytical rates are easier to work with and help to explain the underlying physics.

\section{Results}

\subsection{Slow-light coupled cavity waveguide}

The optical scattering rates (Eq.~\ref{eq7}) indicate that the role of photon reservoir in QD Mollow spectra will be strongest in a media which displays spectral inhomogenety over the width of the Mollow triplets, namely $\omega_L\pm\Omega$. One such relevant example is a PC coupled-cavity waveguide~\cite{Yariv} shown in Fig.~\ref{fig1}(a). Assuming nearest neighbour coupling between adjacent cavities of mode volume $V_{\rm eff}$, an analytical tight binding technique~\cite{Yariv,Fussell1} can be applied to derive an expression for the waveguide photon reservoir function, 
\begin{equation}
J_{\text{ph}}(\omega) 
\!=\!\frac{-{d}^2\omega}{2\hbar\epsilon_0n_b^2 V_{\rm eff}} \frac{1}{\pi}
{\rm Im}\left[ \frac{1}{\sqrt{(\omega-\tilde\omega_u)(\omega-\tilde\omega_l^*)}}\right],
\end{equation}
where $\tilde\omega_{u,l}=\omega_{u,l}\pm i\kappa_{u,l}$\cite{Fussell1},  $\omega_{u,l}$ represents waveguide band-edge frequencies, $\kappa_{u,l}$ represent the damping rates, and $n_b$ is the refractive index of the background dielectric. 
Expressions for photonic LDOS and the reservoir propagator can be derived using $J_{\text{ph}}(\omega)$. The photonic LDOS is proportional to the Purcell factor, PF =$\frac{\gamma}{\gamma_b}$, where $\gamma = {\rm Re}[\int_0^{\infty}d\tau J^D_{\rm ph}(\tau)]$ is the SE rate in the absence of phonon coupling and  $\gamma_b\approx$ 1.5 $\mu$eV is the SE rate of the background slab. The photon propagator is 
\begin{align}
\label{eq19}
\alpha_{\rm P}(\omega) &= \alpha_0(\r1_{\rm D},\r1_{\rm d}) \frac{\omega^2}{4}\abs{\frac{1}{ \sqrt{ (\omega-\tilde\omega_{\rm l}^*) (\omega-\tilde\omega_{\rm u}) }}}^2,
\end{align}
where the prefactor $\alpha_0$ is a frequency-independent geometrical factor. 

\begin{figure}[t]
\vspace{0.cm}
\includegraphics[width=0.99\columnwidth, height=0.99\columnwidth]{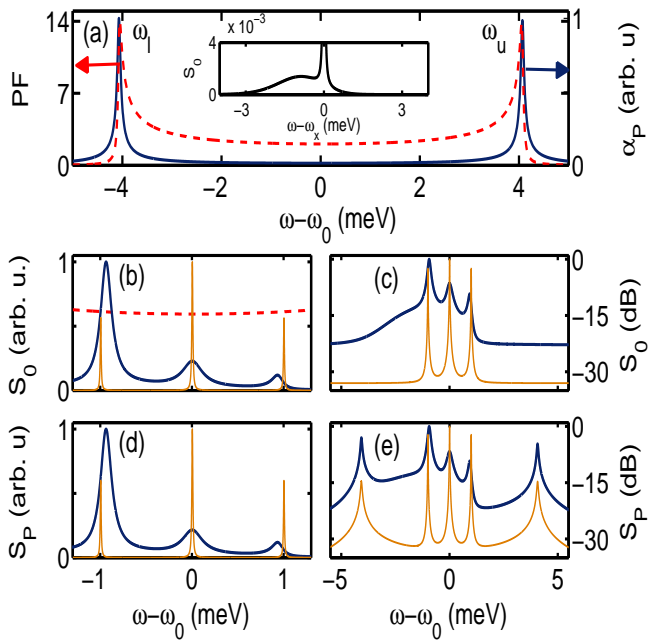}
\vspace{-0.cm}
\caption{\label{fig2}(Color online). \footnotesize{(a) Purcell factor (dashed) and normalized propagator $\alpha_{\rm P}$ (solid) for a coupled cavity PC waveguide. Inset shows normalized emission spectrum from a weakly driven, bare QD. (b) Normalized polarization spectra $S_0$ with (dark) and without (light) phonons near the waveguide band-center (i.e., at $\omega_0$). The dashed line shows the photonic LDOS profile ($\propto$ PF) in normalized units. (c) plots the same quantities as in (b) in a log-scale (dB) over a wider spectral range. (d) Normalized waveguide projected spectra $S_P$, with (dark) and without (light) phonons near the waveguide band-center. (e) plots the same quantities as in (d), in log-scale (dB) and over a wider spectral range to observe mode-edge feeding. The waveguide calculations are similar to those in experiments~\cite{Kuramochi} (see text); the QD dipole moment is taken to be $d = 50$ Debye (0.021 e-nm), T = 4 K, $\gamma_d$ = 7.8 $\mu$eV and  $\gamma_b$ = 1.5 $\mu$eV.}}
\end{figure}

Figure~\ref{fig2} (a) plots the Purcell factor (PF, dashed) and normalized $\alpha_P$ (solid), respectively, for a waveguide formed by coupling individual cavities of $Q$ (quality factor) factor $\approx$ 52000. As shown in Fig.~\ref{fig2}(a), the waveguide structure presents spectral regions, with slow (middle band) and fast (mode-edges, $\omega_{\rm u,l}$) variation of photonic LDOS (Fig.~\ref{fig2}(a)). To develop a simple physical understanding of emitted spectra, we initially consider a QD tuned to the band-center ($\omega_x=\omega_0$) of the waveguide (Fig.~\ref{fig2}(b)). This is a region where the photonic LDOS is almost constant over a broad spectral range ($\approx$ 4 meV, dashed line Fig.~\ref{fig2}(b)) and the polarization spectra $S_0$ of a resonantly-driven QD ($\Delta_{xL} = 0$) without phonons (solid light), resembles a symmetric Mollow triplet with three peaks located at $\omega = \omega_0$, $\omega_0\pm\Omega$, where drive amplitude $\Omega$ = 1 meV. This symmetry of the Mollow sidebands about the drive frequency is expected in the absence of any variations of the photonic LDOS~\cite{Mollow, Ge}. When the effects of phonon interactions are included, strong asymmetries arise amongst the sidebands of the polarization spectrum $S_0$ (dark solid, Fig.~\ref{fig2}(b)). The intensity of the peaks reduce from left to right, suggesting enhanced emission at lower energies. This asymmetry in the Mollow sidebands has been previously observed in phonon-dressed Mollow spectra from bare QDs~\cite{Ulhaq} (i.e., not in a structured reservoir) and it arises due to unequal phonon emission and absorption rates at low temperatures~\cite{Roy,DaraN}. The phonon-induced asymmetry can also be observed in the weak excitation spectra of bare QDs\cite{Axt, Kaushik3} (inset, Fig.~\ref{fig2}(a)), which  resembles the IBM spectrum~\cite{Imamoglu}.  Phonon-induced emission asymmetry of the Mollow spectra has been previously explained using a polaron ME of Lindblad form, derived assuming small drive strengths $\Omega$~\cite{Roy, Roy3}; however, the spectrum calculations using the new phonon scattering terms (Eq.~\ref{eq5}) are valid for large $\Omega$, and produce an even stronger asymmetry between the sidebands that can be attributed to the newly presented phonon scattering term $\Gamma_u$~\cite{Ross} (see later). Apart from the intensity asymmetry, phonons also cause spectral broadening and reduced splitting of the sidebands (dark solid, Fig.~\ref{fig2}(b)). The power dependent broadening also known as excitation induced dephasing arises due to the $\Omega^2$ dependence of the phonon terms (Eq.~\ref{eq5}) and the reduced splitting happens due to renormalization of the drive amplitude (i.e. $\Omega_R\rightarrow\avg{B}\Omega$)~\cite{Roy3}. For example at a phonon bath temperature of 4\,K, then $\avg{B}\approx0.9$, causing a 10\% reduction of the coherent Rabi drive.  Figure~\ref{fig2}(c) plots the polarization spectra $S_0$ over a broader frequency range, in a log scale (dB), which shows the phonon-induced asymmetries more clearly and these will be useful in understanding the role of the waveguide mode-edges in the projected spectra, discussed below.

The reservoir projected spectra $S_P$ (Fig.~\ref{fig2}(d)) is obtained by projecting $S_0$ with $\alpha_P$~\cite{Kaushik3}, accounting for phonon emission from the QD to a detector. The projector function $\alpha_P$ (Fig.~\ref{fig2}(a), dark solid) can be approximated as a sum of two sharp Lorentzians about the waveguide mode-edges and is roughly constant around the region of the Mollow triplet. The projected spectra (Fig.~\ref{fig2}(d)) hence resembles the polarization spectra (Fig.~\ref{fig2}(b)) in this region. The spectral regions near the PC mode-edges are however strongly magnified in the projected spectra (Fig.~\ref{fig2}(e)) compared to the polarization spectra (Fig.~\ref{fig2}(c)) leading to the appearance of  five distinct peaks (Fig.~\ref{fig2}(e)). The spectra is plotted in a log scale (dB) in Fig.~\ref{fig2}(e) to better show the appearance of the five peaks in $S_P$. Phonon induced asymmetry effects also appear in the peaks around the mode-edges (dark line Fig.~\ref{fig2} (e)), which is absent in the non-phonon case (light line).


\begin{figure}[t!]
\vspace{0.cm}
\includegraphics[width=0.98\columnwidth]{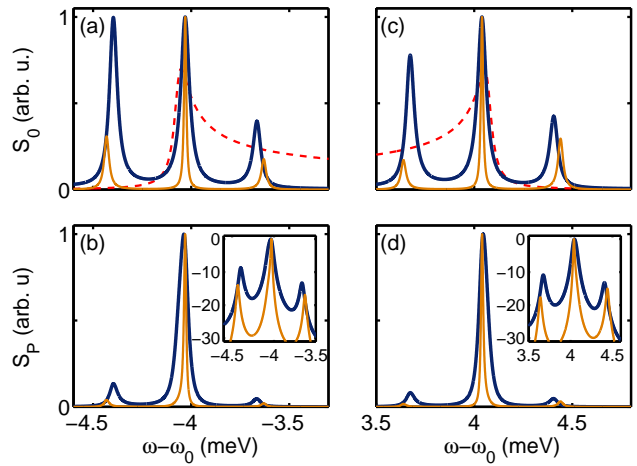}
\vspace{-0.cm}
\caption{\label{fig3}(color online). \footnotesize{Normalized polarization spectra $S_0$ of a QD located near the lower (a) and the upper (c) mode-edge of a waveguide. (b) and (d) show the normalized projected spectra $S_P$ corresponding to (a) and (c). The dark (light) solid lines show calculations with (without) phonons. The insets in (b) and (d) shows $S_P$ in a log scale. The light dashed line shows the normalized photonic LDOS profile, as a reference in (a, c). Calculations use the same parameters as in Fig.~\ref{fig2}.}}
\end{figure}

Near band center, 
the Mollow spectra investigated in Fig.~\ref{fig2} (b) is minimally influenced by the photonic LDOS since it is roughly constant  near the Mollow triplet frequencies (dashed line, Fig.~\ref{fig2}(b)) and photon effects are only observed in the projected spectra near mode-edges, far away from the Mollow peaks. 

To better investigate the role of photon-phonon dynamics in the Mollow spectra, we next calculate the resonant ($\Delta_{xL}$ = 0 meV) spectrum for the QD tuned near the waveguide mode-edge (Fig.~\ref{fig3}). Once again, solid dark and light lines represent calculation with and without phonons. Figures~\ref{fig3} (a) and (b) plot the normalized polarization spectra $S_0$ and projected spectra $S_P$, respectively, for a QD tuned near the lower mode-edge and Fig.~\ref{fig3}(b) and (d) plots the same quantities near the upper mode-edge. Since the bandwidth of the waveguide mode-edges where the photonic LDOS varies maximally is small, we choose a smaller value of $\Omega$ = 0.4 meV for calculating the Mollow spectra. This way, the three Mollow peaks sample different regions of photonic LDOS. For the Mollow spectra near the lower and upper mode-edge without phonons (light line, Fig.~\ref{fig3}(a) and (c)), the intensity of the sideband sampling a lower photonic LDOS (outside waveguide band) is stronger than the one sampling a higher photonic LDOS (inside the waveguide band). Similar intensity asymmetry has been previously reported in case of a Lorentzian cavity~\cite{Tanas}, where the Mollow sidebands sample spectral regions with different photonic LDOS. It is possible to derive the emission spectra analytically in the limit of a Lorentzian cavity and derived spectrum shows that the brighter Mollow sideband samples a smaller photonic LDOS (see dashed line of Fig. 5 in Ref.~\cite{Tanas}). {The current waveguide results shows similar behavior and  here we go beyond previous photon reservoir theories using the secular approximation~\cite{Florescu}, which predicts equal sideband intensities irrespective of asymmetries of the underlying photon bath.} 

When phonons are included, the intensity of the side-bands increase (dark solid lines, Fig.~\ref{fig3} (a, c)). This phonon-induced enhancement outside the waveguide band can limit the efficiency of atomic switching schemes using the mode-edge LDOS asymmetry~\cite{John2}. Since the bandwidth of the waveguide ($\approx$ 8 meV) is greater than that of the phonon bath ($\approx$ 5 meV), emission near band-edge far from the QD is negligible and hence ignored in Fig.~\ref{fig3}. As mentioned above, phonons enhance emission at low energies. Near the upper mode-edge, this leads to a strong enhancement of the Mollow side-band inside the waveguide band compared to the one outside. Thus the upper mode-edge (Fig.~\ref{fig3} (c)) presents an interesting spectral region where the phonon and photon bath act against each other in determining the final intensities of the Mollow side-bands. This is not the case however for the lower mode-edge, where the lower Mollow side-band always remains stronger than the upper one. {Note that previous mean-field treatments~\cite{Roy1} predicting a simple phonon renormalization of the QD-photon bath coupling will not produce such intensity asymmetry between the Mollow sidebands.} The projected spectrum $S_P$ (Fig.~\ref{fig3} (b, d)) show almost an order of magnitude lowering of the Mollow side-bands compared to the central line (at $\omega_x$), due to strong suppression by the narrow Lorentzian projector at the waveguide mode-edge. The corresponding insets plots the respective projected spectra in a log-scale (dB) and shows that the Mollow side-band asymmetries of the polarization spectra $S_0$, arising due to the photon and phonon baths are retained in $S_P$.

\begin{figure}[th!]
\vspace{0.cm}
\includegraphics[width=0.98\columnwidth]{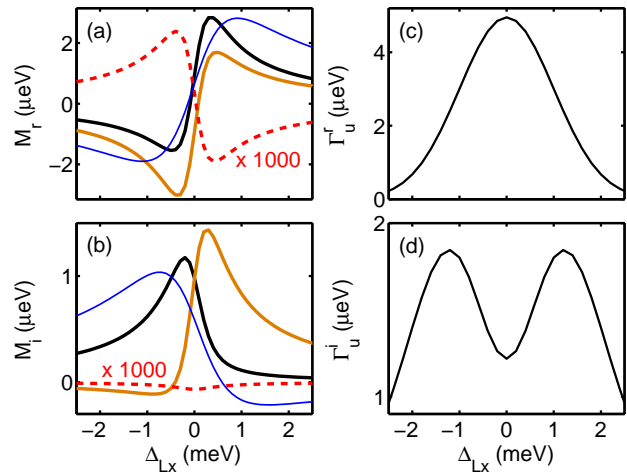}
\vspace{-0.cm}
\caption{\label{fig4}(color online). \footnotesize{ Real (a) and imaginary (b) parts of $M(\Omega)$ as a function of QD-laser detuning $\Delta_{Lx}$, for a QD located near lower mode edge (thick dark), upper mode-edge (thick light) and middle (dashed) of the waveguide band for $\Omega$ = 0.4 meV. The dashed line is multiplied by a factor of 1000 for better visibility. The thin dark line shows calculations for a QD near lower mode-edge at $\Omega$ = 1 meV.  Calculations use the same waveguide parameters as in Fig.~\ref{fig2}. Real (c) and imaginary (d) parts of $\Gamma_u$ as a function of QD-laser detuning $\Delta_{Lx}$ for $\Omega$ = 0.4 meV and T = 4 K. Calculations use the same parameters as Fig.~\ref{fig2}.} }
\end{figure}

At the level of the polaron ME, the incoherent scattering terms $M'(\Omega_R)$ (Eq.~\ref{eq7}) and $\Gamma_u$ (Eq.~\ref{eq5}) are directly responsible for the photon and phonon-induced intensity asymmetries of the Mollow side-bands, respectively. They also represent similar mathematical terms at the level of the ME (Eqs.~\ref{eq5} and \ref{eq7}), namely the commutator of population $\sigma_z$  and polarization $\sm$ and similar terms have been shown previously to be responsible for phonon induced asymmetries of vacuum Rabi doublets in a strongly coupled QD-cavity systems~\cite{Kaushik3, Milde, Ota}. The real and imaginary parts of $M(\Omega) = M_r(\Omega) +iM_i(\Omega) $ ($\Gamma_u = \Gamma^r_u + i\Gamma^i_u$) are plotted in Fig.~\ref{fig4} (a) ((c)) and (b) ((d)), respectively, as a function of detuning $\Delta_{Lx}$, for a fixed drive amplitude $\Omega$. The detuning is varied by tuning the QD with respect to a fixed laser frequency. Though the phonon modified rate, $M'(\Omega_R)$ enters into the ME (Eq.~\ref{eq3}), in this discussion we focus on the behavior of rate $M(\Omega)$, which is not modified by phonons. This does not affect the reasoning below, since for the parameters used, phonons cause very little difference (not shown) in $M(\Omega)$, other than a slight reduction near the optimum points (Fig.~\ref{fig4}(a, b)).  The thick dark, dashed and light line in Fig.~\ref{fig4}(a) plots $M(\Omega)$ when the drive laser is located near lower mode-edge, band-center and upper mode-edge of the waveguide band respectively. As is the case with most of our Mollow spectra calculations, the three plots correspond to a value of $\Omega$ = 0.4 meV. Calculations for higher drives (thin dark line, $\Omega$ = 1 meV) show similar behavior to calculations for lower drives (thick dark line, $\Omega$ = 0.4 meV), when the laser is located near the lower mode-edge. The phonon term, $\Gamma_u$ is not influenced by photon bath and is plotted for $\Omega$ = 0.4 meV (Fig.~\ref{fig4}(c, d)). Note that we have  verified analytically and numerically (not shown) that only the real parts of $M(\Omega)$ and $\Gamma_u$ contribute to any intensity asymmetries of the Mollow spectra. As expected from Fig.~\ref{fig2}(b), $M(\Omega)$ is smaller by three orders of magnitude at the band-center compared to the mode-edges and the photonic LDOS does not influence the Mollow spectra (solid light line, Fig.~\ref{fig2}(b)). 

The full resonant Mollow spectra with phonons (dark solid line) from Fig.~\ref{fig2}(b) suggest that a positive value of $\Gamma^r_u$ (at $\Delta_{xL}$ = 0 meV) correlates with a strong lower Mollow side-band compared to the upper one. The corresponding resonant $M_r(\Omega)$ near the lower mode-edge (Fig.~\ref{fig4}(a), thick dark solid, at $\Delta_{xL} = 0$ meV) has a positive value (0.93) and produces a similar asymmetry in the spectrum without phonons (light line, Fig.~\ref{fig3}(a)). A negative resonant $M_r(\Omega)$ ($=-0.95$) near the upper mode-edge (Fig.~\ref{fig4}(a), thick light solid) leads to opposite assymetry of Mollow sidebands, without phonons, (light line, Fig.~\ref{fig3}(c)). Since $\Gamma^r_u\ge M_r(\Omega)$ at $\Delta_{xL} = 0$ meV, the sideband of the final spectrum (thick dark, Fig.~\ref{fig3}(c)) is stronger to the left.

\begin{figure}[th!]
\vspace{0.cm}
\includegraphics[width=1\columnwidth]{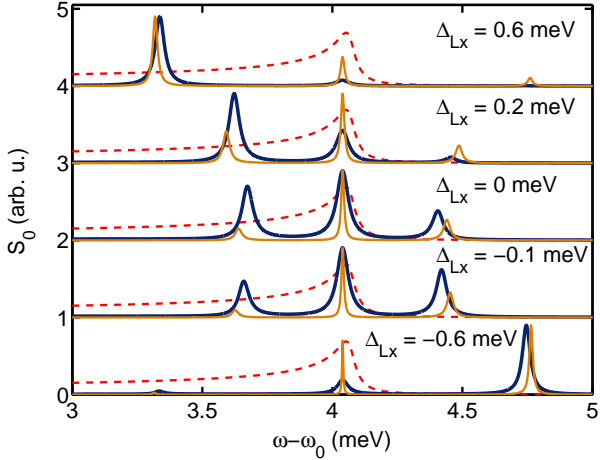}
\vspace{-0.cm}
\caption{\label{fig5}(color online). \footnotesize{ Polarization spectra $S_0$ of a QD with (dark) and without (light) phonon coupling, plotted for different QD-laser detunings $\Delta_{Lx}$. The dashed line shows the normalized profile of the photonic LDOS and the laser is spectrally fixed (central Mollow line) near the upper waveguide mode-edge, with drive strength $\Omega$ = 0.4 meV. Calculations use the same parameters as in Fig.~\ref{fig2}.}}
\end{figure}

The role of $M_r(\Omega)$ term in governing the asymmetry of the Mollow side-bands becomes more clear, when the spectrum is plotted as a function of detuning $\Delta_{Lx}$. Figure~\ref{fig5} plots the normalized polarization spectrum with (thick dark solid) and without phonons (thin light), for different detunings $\Delta_{Lx}$, near the upper mode-edge of the waveguide (dark dashed). For small negative detunings ($\Delta_{Lx} = -0.1$ meV), $M_r(\Omega)$ has a larger negative value (-2) compared to its value ($-0.95$) at $\Delta_{Lx} = 0$ meV (thick light, Fig.~\ref{fig4}(a)), and the spectrum without phonons (thin light) shows a larger asymmetry between Mollow side-bands, with a strong upper band. Inclusion of phonons (thick dark) reduces this assymetry since $\Gamma^r_u > \abs{M_r(\Omega)}$. For a small positive detuning ($\Delta_{Lx}$ = 0.2 meV), $M_r(\Omega)$ is positive (= 1) (thick light, Fig.~\ref{fig4}(a)) and the lower Mollow side-band without phonons is now stronger (light line, Fig.~\ref{fig5}). A larger value of $\Delta_{Lx}$ is chosen for the positive detuning case since at $\Delta_{Lx}$ = 0.1 meV, $M_r(\Omega)$ is very small ($=0.2$).  When detuning $\Delta_{Lx}$ is large compared to drive amplitude $\Omega_R$ (= 0.4 meV), the side-band spectrally closer to the QD dominates. This is verified by spectrum calculations at large detunings ($\Delta_{Lx}$ $\pm$ 0.6 meV, top and bottom curves, Fig.~\ref{fig5}), where the sideband closer to the QD is strongest and the importance of the assymetry terms ($M_r(\Omega)$, $\Gamma^r_u$) becomes minimal.

\subsection{Disordered W1 photonic crystal waveguide}

\begin{figure}[t!]
  \centering
  \includegraphics[scale=0.9]{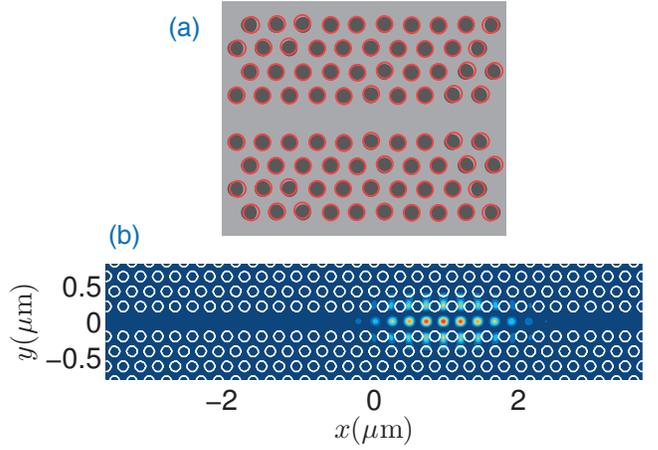}
  \caption{\label{fig6} (Color Online) \footnotesize{ (a) Top view schematic of a
  disordered W1 PC\ slab with ideal air holes (dark-filled/black circles),
  disordered holes (light/red circles) and the background slab
  (gray). 
    (b) The mode profile $|E_y|^2$, at slab center, along with the numerical dielectric profile
    for a disordered fundamental mode at the disorder-induced resonance $\omega_{c}$ (see also Fig.~\ref{fig7}(a) for the corresponding LDOS profile.}).}
\end{figure}

The role of the interacting photon and phonon baths on the QD spectra were explored so far in the case of a model coupled-cavity waveguide. Since the theory is valid for arbitrary structured reservoirs, we can apply it to explore such dynamics in any desired LDOS reservoir function.
By way of an example,  we  next consider a W1  PC waveguide (cf.~Fig.~\ref{fig1} (b)) which is formed by introducing a line
defect in a triangular lattice of air holes etched in a semiconductor
slab. To further demonstrate the strength of out technique, we also add in realistic structural disorder to the waveguide, using full
3D numerical simulations. The disorder perturbation is characterized by a random hole center
shift drawn from a normal probability distribution having a zero mean and non-zero 
standard deviation denoted by $\sigma$. This is shown schematically in
Fig.~\ref{fig6} (a). We choose
a pitch of $a=240\,\mathrm{nm}$, and model GaAs membranes
with a slab dielectric constant of $\varepsilon = 12.11$. In
terms of $a$, we use a corresponding W1 waveguide with
a hole radius of $r=0.295a$, slab height of $h=0.625a$,  and the length of the waveguide is fixed at $30a$. 
Intrinsic disorder is present due to manufacturing imperfections,
and is chosen to be $\sigma = 0.005a=1.2\,$ nm, which has been shown to match
 the DOS of related experiments~\cite{Garcia}. With these parameters, a $y$-oriented dipole
placed at the center of the waveguide (an anti-node for the field component
$E_{y}$) excites the fundamental
mode modified by the intrinsic disorder present in the waveguide.
The resulting Purcell factor calculated from a full 3D FDTD simulation is shown
in Fig.~\ref{fig6}(a) and the disordered electric-field mode profile ($|E_{y}|^{2}$) at $\omega_{c}$ (center of the mode resonance, formed near the W1 mode edge) is 
shown Fig.~\ref{fig6}(b), where we show the field profile at slab center; see Ref.~\onlinecite{Mann} for  numerical implementation details.

\begin{figure}[h!]
\vspace{0.cm}
\includegraphics[width=1\columnwidth]{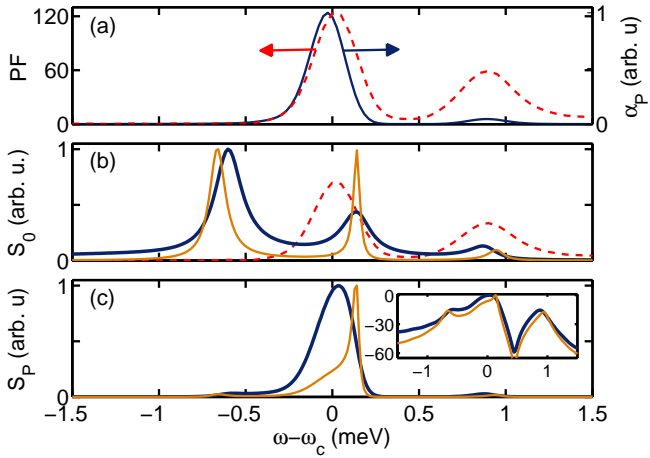}
\vspace{-0.cm}
\caption{\label{fig7}(color online). \footnotesize{ (a) Purcell factor (dashed) and normalized propagator $\alpha_P$ for a disordered W1 PC waveguide. $\omega_c$ marks the band-edge where the Purcell factor is maximum. Normalized polarization spectra $S_0$ (b) and normalized projected spectrum $S_P$ (c) for a QD, with (dark) and without (light) phonon coupling, near the waveguide mode edge. The dashed line in (b) shows the photonic LDOS profile ($\propto$ PF) in normalized units. Inset of (c) shows the projected spectra on a log scale. The QD parameters are same as in Fig.~\ref{fig2}, $\Delta_{Lx}$ = 0 meV, and the waveguide calculations use parameters from Ref.~\onlinecite{Mann}. }}
\end{figure}

Using the above W1 photon bath LDOS results, 
Fig.~\ref{fig7}(a) plots the Purcell factor and normalized projector for the disordered W1 waveguide at the field antinode position. Both the PF and the projector show two resonances close to the mode-edge. Figure~\ref{fig7}(b) plots the polarization Mollow spectra for a QD tuned near the mode-edge (dashed line) and the dark (light) solid lines show on-resonant Mollow spectra, with (without) phonons. The drive amplitude $\Omega_R$ is chosen to closely match the spectral separation between the two LDOS resonances. As expected from Fig.~\ref{fig3}(a), phonons and photons strongly enhance the intensity of the lower side-band, sampling a smaller photonic LDOS. Figure~\ref{fig7}(c) plots the projected spectra in linear (main) and log-scale (inset) and the side-bands are almost invisible in the linear spectra, due to strong suppression by the projector (Fig.~\ref{fig7}(a), dark line).

\section{Conclusions}
 We have investigated  the resonance florescence spectra from a strongly driven QD coupled to  slow-light PC waveguides, and presented a theory that allows one to include both photon and phonon coupling
in the presence of a strong pump field. A strong spectral variation of the coupled photon and phonon LDOS values cause a rich intensity asymmetry in the Mollow triplet sidebands from a coherently driven QD. As a first  example of a structured photon reservoir, we  chose a coupled cavity PC waveguide which shows spectral regions that support both fast and slow spectral variations of the photonic LDOS. Spectral inhomogeneities of the photon and phonon baths separately influence the side-band intensities and their separate roles are  identified by investigating the Mollow spectra near the upper mode-edge of the coupled cavity waveguide. In this slow-light region, photon and phonon baths counteract each other to determine the final side-band intensities.
The important incoherent photon and phonon scattering terms responsible for causing the intensity asymmetry are identified at the level of the polaron ME, where phonon coupling effects are included nonperturbatively.
 The polaron ME technique developed is quite general and can be applied to arbitrary structured photon baths profiles.  As a 
further example, we also studied the case of a driven QD
 in a disordered W1 slab waveguide, and once again showed the important 
role of coupled photon and phonon interactions in the light emission
from a coherent drive.
These dynamical effects of coupling to both phonon and photon baths are  relevant for   understanding related experiments and emerging applications of  QD waveguide systems.

\acknowledgments
This work was supported by the Natural Sciences and Engineering Research Council of Canada. We thank Rongchun Ge for useful discussions.

\end{document}